# Regulatory Migration to Europe:

# ICO Reallocation Following U.S. Securities Enforcement

Krishna Sharma[1], Khemraj Bhatt[2], Indra Giri[3]

**Abstract**

This paper examines whether a major U.S. regulatory clarification coincided with cross-border spillovers in crypto-asset entrepreneurial finance. We study the Securities and Exchange Commission's July 2017 DAO Report, which clarified the application of U.S. securities law to many initial coin offerings (ICOs), and analyze how global issuance activity adjusted across regions. Using a comprehensive global dataset of ICOs from 2014 to 2021, we construct a region-month panel and evaluate issuance dynamics around the announcement. We document a substantial and persistent reallocation of ICO activity toward Europe following the DAO Report. In panel regressions with region and month fixed effects, Europe experiences an average post-2017 increase of approximately 14 additional ICOs per region-month relative to other regions, net of global market cycles. The results are consistent with cross-border regulatory spillovers in highly mobile digital-asset markets.

[1] Hoover Institution, Stanford University, CA
[2] First Citizen Bank, NC
[3] Kathmandu University, Kathmandu

# 1  Introduction

The rapid expansion of crypto-asset fundraising between 2016 and 2018 represents one of the most consequential episodes of entrepreneurial finance in recent decades. Initial coin offerings (ICOs) enabled early-stage ventures to raise capital by selling blockchain-based tokens to a global investor base, often outside traditional financial intermediation and with limited regulatory oversight (1; 20). At their peak, ICOs raised tens of billions of U.S. dollars worldwide and attracted issuers from a wide range of jurisdictions. A growing literature has examined token pricing, disclosure incentives, governance design, and investor behavior in these markets (24; 7; 27; 14; 26). Despite this progress, relatively little is known about how the geographic allocation of ICO issuance responds to cross-border differences in regulatory treatment.

A defining feature of token-based finance is its high mobility of issuance activity. ICOs are globally accessible, rely minimally on domestic intermediaries, and can be launched with limited physical relocation costs. These characteristics suggest that regulatory interventions in one jurisdiction may generate spillovers by reallocating issuance activity toward alternative regions rather than curtailing fundraising altogether. While this intuition features prominently in policy discussions, systematic empirical evidence on such cross-border reallocation remains limited.

A salient regulatory inflection point occurred in July 2017, when the U.S. Securities and Exchange Commission (SEC) released its *Report of Investigation* concerning the Decentralized Autonomous Organization (DAO) (29). The report concluded that DAO tokens constituted securities under the Howey test and articulated reasoning that could apply to many ICOs circulating at the time. Although the report did not introduce new statutes, it marked a clear shift in enforcement expectations and was followed by a sequence of enforcement actions targeting unregistered offerings and misleading disclosures (13; 24).

These developments may have increased regulatory uncertainty and expected compliance costs for U.S.-facing issuers.

This paper examines whether the DAO Report coincided with a reallocation of ICO issuance toward Europe, a region that offered heterogeneous regulatory environments during the ICO boom. Unlike the United States, Europe did not experience a sudden, binding regulatory intervention in mid-2017. Instead, several European jurisdictions such as Switzerland, Estonia, and Malta provided early guidance or innovation-friendly frameworks, while EU level institutions issued non-binding warnings rather than enforceable rules (31; 17). This institutional asymmetry creates a natural setting to assess whether heightened regulatory scrutiny in a dominant jurisdiction is associated with shifts in the geographic distribution of token issuance.

Our empirical analysis uses a global ICO dataset spanning 2014 to 2021 and focuses on region-month variation in issuance activity. Consistent with recent work on crypto-asset markets, we use the DAO Report as a salient regulatory event to document the timing of structural breaks, relative deviations, and reallocation patterns across regions (6; 4). We combine descriptive evidence, region-month panel regressions with fixed effects, and event study specifications to assess whether Europe experienced a distinct post-2017 deviation relative to other regions, net of global ICO cycles. Our objective is to document the timing and magnitude of geographic reallocation patterns rather than to establish tightly identified causal effects of regulatory policy. We find that a substantial and persistent relative shift in ICO issuance toward Europe after July 2017. Europe's monthly ICO counts increase sharply relative to other regions in the same months. These patterns are robust to alternative covariance estimators, exclusion of major European crypto hubs, and leave-one-region-out sensitivity checks.

This paper contributes to three strands of research. First, it adds a geographic and regulatory dimension to the literature on ICO economics by documenting how issuers may adjust location choices in response to enforcement-related uncertainty. Second, it extends the literature on regulatory spillovers and cross-border capital reallocation by showing that digital asset entrepreneurial finance can exhibit rapid geographic mobility following asymmetric regulatory signals (23). Third, the findings inform ongoing policy debates on crypto-asset regulation by illustrating how unilateral enforcement actions may reshape the global distribution of activity, with implications for coordination challenges under emerging frameworks such as the EU's Markets in Crypto-Assets regulation (16).

The remainder of the paper proceeds as follows. Section 2 outlines the institutional and regulatory background in the United States and Europe. Section 3 reviews the related literature. Section 4 describes the data. Section 5 presents the empirical strategy. Section 6 reports the results. Section 8 discusses interpretation, alternative explanations, and policy implications, and Section 9 concludes.

## 2  Institutional and Regulatory Background

Understanding whether the SEC's 2017 DAO Report coincided with cross-border spillovers requires a clear description of the regulatory environments governing ICO activity in the United States and Europe. Although ICOs were marketed globally, issuers remained subject to the jurisdictional rules governing the location of the offering and the targeted investor base. Prior to mid-2017, regulatory approaches to token issuance were fragmented, lightly enforced, and often ambiguous (22; 3; 25). The DAO Report created a sharp regulatory clarification in the United States, whereas Europe maintained heterogeneous environments that were, in many cases, comparatively permissive or provided greater regulatory clarity. This asymmetry is central to interpreting the subsequent patterns in ICO activity.

## 2.1 Regulatory Conditions in the United States and the DAO Report

Prior to July 2017, the SEC had not explicitly applied U.S. securities law to ICOs, though the legal framework existed. While the Howey test provided a legal standard under which many ICOs could potentially be classified as securities offerings, issuers operated in an environment where formal guidance and enforcement signals were still developing (13). Many issuers marketed utility tokens under the assumption that they could avoid classification as securities, and some explicitly excluded U.S. investors as a precaution.

The SEC's *Report of Investigation Pursuant to Section 21(a)* concerning the Decentralized Autonomous Organization marked a turning point (29). The report concluded that DAO tokens were investment contracts and therefore securities, and it articulated reasoning that extended to many ICOs circulating at the time. The announcement signaled a credible enforcement shift and was followed by a series of actions targeting unregistered offerings and misleading disclosures (24; 13). These developments plausibly increased expected compliance costs for U.S.-facing issuers and may have created incentives to relocate offerings to jurisdictions perceived as more permissive or offering greater regulatory clarity.

## 2.2 European Regulatory Conditions During the ICO Boom

Europe did not adopt a unified regulatory framework for ICOs during the 2016 to 2018 period. Instead, the region exhibited a diverse set of national approaches, ranging from explicit guidance to relative regulatory silence. Switzerland's financial regulator, FINMA, published an early taxonomy that differentiated payment, utility, and asset tokens and issued case-by-case assessments for prospective issuers (19). The FINMA guidance explicitly stated: "ICOs are not per se subject to financial market law. However, depending on how they are structured, ICOs may fall under financial market regulations". Estonia pursued a digital innovation strategy supported by its e-residency program. Malta, Gibraltar, and Liechtenstein introduced comprehensive statutory frameworks intended to attract blockchain ventures (31).

At the European Union level, institutions such as the European Securities and Markets Authority and the European Banking Authority released investor warnings highlighting operational and investor-protection risks (17). These statements did not create binding rules and therefore did not materially constrain issuers, though national securities and financial regulations remained applicable depending on jurisdiction. Surveys of the global crypto asset landscape document significant heterogeneity in regulatory treatment across European jurisdictions during this period (9). As a result, many European jurisdictions remained comparatively permissive or provided clearer frameworks relative to the post-DAO U.S. regime.

## 2.3  Potential for Cross-Border Spillovers

The regulatory asymmetry between the United States and Europe may have created incentives for issuers to adjust the geographic location of token sales after July 2017. ICOs were globally accessible and mobile by design, so relocation costs were potentially low. Firms facing heightened enforcement risk in the United States could credibly shift operations to jurisdictions in Europe that offered clearer, more innovation-friendly, or less stringent regulatory environments.

This mechanism aligns with broader evidence on regulatory arbitrage and cross-border capital flows, which shows that firms can reallocate activity when regulatory requirements diverge across jurisdictions (23). If the DAO Report imposed a meaningful deterrent effect on U.S.-oriented ICO issuance, an observable redirection of activity toward Europe would be a natural consequence. The absence of a simultaneous Europe-wide regulatory shock strengthens the plausibility of interpreting any discontinuities around July 2017 as spillovers associated with heightened U.S. enforcement expectations, though alternative explanations must also be considered.

## 3   Related Literature

This paper connects to three strands of research: the economics of initial coin offerings, regulatory spillovers and cross-border reallocation of financial activity, and the evolving literature on crypto-asset regulation and policy.

The first strand examines initial coin offerings as a form of entrepreneurial finance. Early work documents that blockchain ventures share structural similarities with traditional early stage financing while operating under distinct informational frictions (2). Disclosure quality and whitepaper content are shown to be important predictors of investor demand and fundraising success (20), while governance and commitment signals embedded in token design affect outcomes and credibility (21). ICO pricing dynamics and investor returns are studied by Benedetti & Kostovetsky (7), and Momtaz (27) highlights the volatility of ICO markets and their sensitivity to adverse industry events, including regulatory interventions. Formal models of token-based platform finance emphasize the role of token design and incentive alignment (14; 26). Information intermediaries in the ICO ecosystem are analyzed by Boreiko & Vidusso (10). While this literature provides rich evidence on valuation, disclosure, and investor behavior, it offers relatively limited insight into how the geographic allocation of ICO issuance responds to regulatory shocks across jurisdictions.

A second and closely related strand studies regulatory spillovers and the cross-border reallocation of financial activity. In traditional financial markets, differences in regulatory regimes are known to redirect capital flows and business activity internationally (23). In the ICO context, Bellavitis, Cumming & Vanacker (6) provide direct evidence that regulatory interventions can displace token-offering activity across countries. Using a global ICO sample, they document that restrictive actions are associated with short-run increases in issuance elsewhere, accompanied by compositional changes and longer-run market adjustment. Related work shows that regulatory news is a key driver of crypto-market reactions (4) and that legal frameworks shape trading and issuance outcomes in digital-asset markets (18). This literature motivates our focus on whether a prominent U.S. enforcement signal, the SEC's 2017 DAO Report, coincides with a reallocation of ICO activity toward alternative jurisdictions rather than an overall market contraction.

A third strand examines crypto-asset regulation and the institutional challenges posed by globally integrated digital markets. Legal and policy scholarship documents the fragmentation of national regulatory approaches and the coordination problems that arise in the absence of harmonized standards (3). Comparative analyses highlight substantial heterogeneity in how jurisdictions classify and regulate tokens (31), while doctrinal work examines the application of securities law to token sales and enforcement-based regulatory strategies (22; 25; 13). The global regulatory landscape is surveyed by Blandin et al. (9), with broader policy perspectives provided by European Central Bank, 2022 (16) and Auer et al. (5). Legal scholarship on blockchain governance is reviewed in Wright & De Filippi (30) and Chen (12). Although this literature clarifies institutional environments and regulatory intent, it provides limited empirical evidence on how issuers respond to enforcement signals through cross-border relocation.

## 4  Data

The empirical analysis draws on a global project-level dataset of initial coin offerings (ICOs) assembled from multiple publicly available archival sources. The objective of the data construction process is to obtain a harmonized and internally consistent record of token issuance activity across regions and over time, rather than to recover a complete census of all historical offerings. Given the fragmented disclosure environment of early ICO markets, particular care is taken to standardize issuer location, launch timing, and basic project characteristics.

## 4.1 Data Sources and Coverage

The core data source is the Token Offerings Research Database (TORD) compiled by Momtaz (28), which provides project-level identifiers for ICOs, initial exchange offerings (IEOs), and security token offerings (STOs) through 2021. To improve coverage prior to the 2017 to 2018 market peak and to cross-validate key fields, TORD is supplemented with historical records from multiple ICO-tracking platforms (including ICOBench, ICODrops, TokenData, and CoinSchedule), archived project whitepapers, press releases, and blockchain industry repositories.

All project entries are cross-checked across at least two independent archival sources whenever possible. Approximately 78% of projects in the final sample are verified across multiple sources. Projects with inconsistent or unverifiable information regarding launch timing or issuer location are excluded. This conservative inclusion criterion prioritizes internal consistency and comparability over maximal coverage, particularly in the early years of the sample.

## 4.2 Geographical Classification

Issuer location is standardized using a hierarchical classification procedure. Each project is assigned to a country based on the registered legal entity location when available. If registration information is unavailable, issuer headquarters as stated in the whitepaper or official project documentation is used. In cases where multiple locations are reported, priority is given to the registered entity location, followed by the primary operational headquarters. Approximately 84% of projects have explicitly documented headquarters or registration information.

Countries are aggregated to the regional (continental) level for the main analysis. The eight regions in the analysis are: Europe, Asia, North America, South America, Africa, Oceania, Middle East, and Other. Europe is defined to include EU member states as well as closely integrated European jurisdictions commonly active in ICO markets (including Switzerland, the United Kingdom, Norway, Liechtenstein, and Gibraltar). Projects are assigned to Europe if the issuer location falls within one of these jurisdictions. Alternative regional groupings and exclusions of specific crypto-friendly jurisdictions are explored as robustness checks. A complete list of countries classified as European is provided in Appendix Table A2.

## 4.3 Temporal Aggregation and Panel Construction

Token launch dates are standardized to monthly frequency. When exact sale start dates are reported (approximately 71% of projects), the project is assigned to the corresponding calendar month. When only partial timing information is available, the earliest verifiable public sale or announcement date is used. All months between January 2014 and December 2021 are included in the panel.

The primary unit of observation is a region-month. For each region and month, we compute the number of ICOs launched, the total amount of funds raised in U.S. dollars, and Europe's share of global ICO activity measured both by the count of offerings and by the volume of capital raised. These measures jointly capture the intensity, scale, and geographic allocation of ICO financing over time. An analogous balanced monthly aggregation is constructed for the global market to facilitate relative comparisons.

## 4.4 Fundraising Amounts

Fundraising amounts are recorded in nominal U.S. dollars without inflation adjustment, as the analysis period is relatively short (2014 to 2021) and the primary focus is on cross-region comparisons within the same months. For offerings denominated in cryptocurrencies (e.g.,

ETH or BTC), reported amounts are converted to USD using contemporaneous exchange rates at the time of the token sale. Because fundraising data are highly right-skewed (the top 5% of offerings account for approximately 65% of total capital raised) and disproportionately influenced by a small number of very large offerings, and because fundraising amounts are missing for approximately 38% of projects, fundraising outcomes are treated as secondary and are primarily used in robustness and validation exercises rather than as the main outcome for inference.

## 4.5 Final Sample and Summary Statistics

The final analysis sample consists of a balanced monthly panel covering 96 months (January 2014 to December 2021) and eight major regions, yielding 768 region-month observations. Summary statistics are reported in Section 6. Additional details on data construction, source reconciliation, and variable definitions are provided in Table A1 of Appendix A.

# 5 Empirical Strategy

This section describes the empirical framework used to assess whether the SEC's July 2017 DAO Report is associated with a relative reallocation of ICO activity toward Europe. Consistent with recent empirical work on crypto-asset markets and regulatory event studies (6; 24; 4), we use the DAO Report as a salient regulatory timing anchor to document structural breaks, relative deviations, and reallocation patterns in ICO issuance across regions. Our objective is to characterize the timing and magnitude of geographic reallocation patterns rather than to establish tightly identified causal effects of regulatory policy.

The empirical strategy combines descriptive pre-post comparisons, cross-region panel regressions with time fixed effects, and dynamic event-study specifications. Together, these approaches allow us to evaluate whether Europe experienced a distinct post-2017 increase in ICO activity relative to other regions, net of global market cycles and common shocks.

However, the analysis faces important limitations. The small number of regional units (eight continents) constrains statistical power and limits the feasible range of inference approaches, particularly regarding clustering choices. Moreover, the observational nature of the data and the presence of potential confounding factors preclude strong causal claims.

## 5.1 Outcomes and Unit of Observation

The unit of observation is a region-month. Let $r$ index regions (continents) and $t$ index months. Our primary outcome variable, $Y_{rt}$, is the number of ICOs launched in region $r$ during month $t$. We focus on ICO counts as the main outcome because they are consistently observed across regions and over time and are not subject to the substantial missingness that affects fundraising amounts.

## 5.2 Baseline Relative-Comparison Specification

A key challenge in this setting is that the ICO market experienced a pronounced global boom during 2017 to 2018. Simple Europe-only before/after comparisons risk conflating a Europe-specific response with worldwide market expansion. To isolate Europe's differential deviation from global ICO cycles, our preferred specification estimates a relative-comparison panel model with region and month fixed effects:

$$Y_{rt} = \alpha_r + \gamma_t + \beta\,(\text{Europe}_r \times \text{Post}_t) + \varepsilon_{rt}, \tag{1}$$

where $\alpha_r$ are region fixed effects, $\gamma_t$ are month fixed effects, $\text{Europe}_r$ is an indicator equal to one for Europe, and $\text{Post}_t$ equals one for months on or after July 2017. The coefficient $\beta$ captures the average differential change in European ICO activity after the DAO Report relative to other regions in the same months.

Month fixed effects absorb global shocks and common market dynamics affecting all regions simultaneously, including the overall ICO boom and subsequent market contraction. Region fixed effects absorb time-invariant differences across regions, such as baseline financial development, fintech ecosystems, and reporting intensity. We interpret a positive and stable estimate of $\beta$ as evidence consistent with a post-2017 reallocation of ICO issuance toward Europe, though we cannot rule out other concurrent explanations.

This specification follows standard practice in empirical studies that document market responses to regulatory interventions without imposing a formal causal policy-evaluation framework (6; 4).

## 5.3 Inference and Small-Sample Considerations

The panel comprises N = 8 regions observed over T = 96 months, which raises small-sample concerns for conventional cluster-robust inference. Clustering standard errors by region is potentially unreliable with so few clusters, as asymptotic cluster approximations perform poorly when the number of clusters is small (8; 11). Our baseline tables therefore report month-clustered standard errors.

However, month clustering alone does not guarantee robustness to serial correlation within regions over time. To strengthen inference under conservative dependence assumptions, we complement month clustering with two approaches that are well-suited to panels with large T and small N: (i) two-way clustering by region and month, and (ii) Driscoll–Kraay standard errors, which are robust to general forms of temporal dependence and cross-sectional dependence in panels. We report these alternative estimators as key robustness checks and interpret results primarily through the consistency of estimates across inference methods rather than any single asymptotic approximation.

Finally, we also use placebo treated-region exercises to assess whether Europe's post-2017 break is exceptional relative to other regions, consistent with a descriptive "randomization style" validation in small-N settings.

## 5.4 Event-Study Evidence

To examine the timing, persistence, and potential anticipation of the post-2017 reallocation, we estimate an event-study version of equation (1). Let $k = t-t_0$ denote event time, where $t_0$ corresponds to July 2017. For each integer $k$ in the event window $k \in [-12,+24]$, define an indicator $1\{t - t_0 = k\}$. The event-study specification is:

$$Y_{rt} = \alpha_r + \gamma_t + \sum_{k \in \mathcal{K},\, k \neq -1} \beta_k \left(\text{Europe}_r \times 1\{t - t_0 = k\}\right) + \varepsilon_{rt}, \quad (2)$$

where $k = -1$ (the month prior to the DAO Report) is omitted and serves as the reference category. The coefficients $\{\beta_k\}$ trace Europe's differential ICO activity relative to other regions as a function of event time.

This event-study design is used as a diagnostic and validation tool. We assess whether pre-event coefficients exhibit systematic anticipatory divergence and whether post-event coefficients display a discrete and persistent shift. Evidence of flat pre-period coefficients followed by sustained positive post-period coefficients is consistent with a reallocation response occurring after the regulatory announcement rather than a gradual secular trend. However, even in the absence of pre-trends, we cannot definitively attribute the post-period increase to the DAO Report alone, as other contemporaneous factors could also explain the pattern.

# 6 Results

## 6.1 Descriptive Patterns

We begin with descriptive evidence on the evolution of ICO activity across regions. Figure 1 plots monthly ICO counts in Europe and the rest of the world using a three-month moving average to smooth short-run volatility. The series exhibits a pronounced global expansion during 2017 to 2018 followed by a sharp contraction, consistent with the documented ICO boom-bust cycle. Visually, Europe's trajectory appears to diverge upward relative to the rest of the world beginning around mid-2017.

Table 1 reports summary statistics for Europe and the rest of the world before and after the DAO Report (July 2017). Europe accounts for a substantially larger volume of ICO activity in the post-2017 period both in absolute terms and as a share of global issuance. In the pre-DAO period, Europe launched 109 ICOs compared to 145 in the rest of the world. In the post-DAO period, these numbers increased to 1,008 and 1,014 respectively, indicating that Europe's growth rate substantially exceeded that of other regions.

Table 1: Summary Statistics by Period and Region

| Region | Period | Total ICOs | Total Funds ($B) | Median Funds ($M) | Obs (Region-Months) |
|---|---|---|---|---|---|
| Europe | Pre-DAO | 109 | 0.50 | 5.20 | 42 |
| Rest of World | Pre-DAO | 145 | 0.98 | 2.75 | 294 |
| Europe | Post-DAO | 1,008 | 4.64 | 3.48 | 54 |
| Rest of World | Post-DAO | 1,014 | 5.12 | 4.75 | 378 |

*Notes:* Pre-DAO period is January 2014 to June 2017. Post-DAO period is July 2017 to December 2021. Fundraising statistics are computed for the subset of ICOs with non-missing capital raised (approximately 62% of projects).

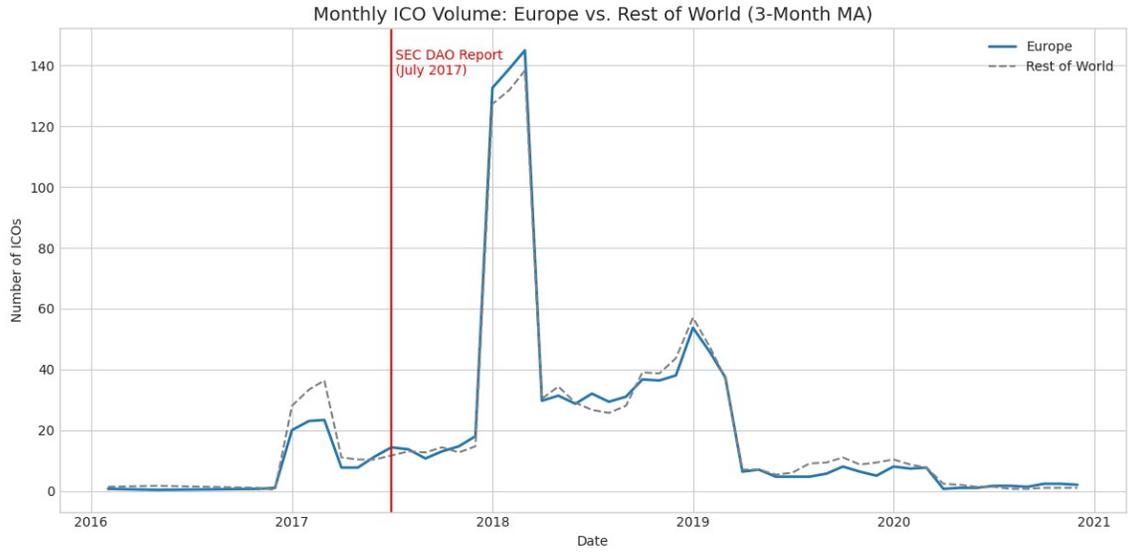

Figure 1: Monthly ICO counts in Europe and the rest of the world (3-month moving average). The vertical line marks the SEC DAO Report (July 2017).

Figure 2 plots Europe's share of global ICO counts over time. Europe's share rises noticeably following mid-2017, motivating a formal region-month regression framework that absorbs common global shocks through month fixed effects.

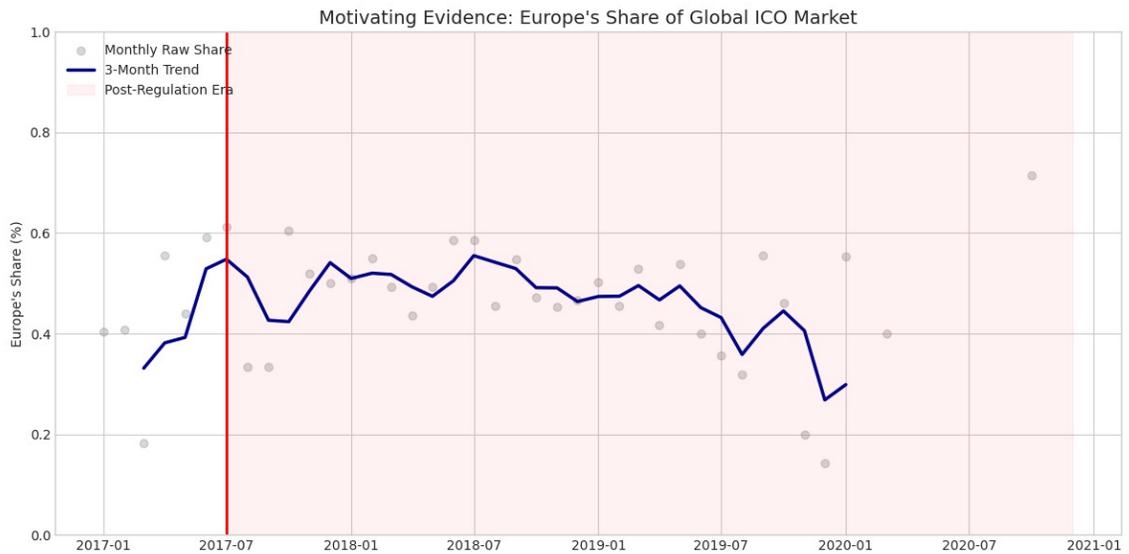

Figure 2: Europe's share of global ICO activity (count-based). The increase in Europe's share after mid-2017 is apparent, though subject to high-frequency volatility.

## 6.2 Relative Reallocation Toward Europe

We next estimate a region-month panel regression with region and month fixed effects, where identification comes from Europe's differential deviation after July 2017 relative to other regions in the same months. Table 2 reports the baseline estimate.

Table 2: Relative Reallocation of ICO Activity Toward Europe

|  | ICO Count |
|---|---|
| Europe × Post | 13.88** |
|  | (6.63) |
| Pre-treatment mean (Europe) | 2.60 |
| Region Fixed Effects | Yes |
| Month Fixed Effects | Yes |
| Observations | 768 |
| Regions | 8 |
| R-squared | 0.847 |

*Notes:* The dependent variable is the number of ICOs launched in a region-month. The coefficient represents Europe's average differential increase in ICO activity relative to other regions after July 2017, controlling for region and month fixed effects. Standard errors clustered by month (96 clusters) are reported in parentheses. The pre-treatment mean is the average number of ICOs per month in Europe during January 2014 to June 2017. ** $p < 0.05$, * $p < 0.10$.

The estimate indicates a substantial increase in European ICO activity relative to other regions after the DAO Report, net of global market cycles absorbed by month fixed effects.

Specifically, Europe experiences an average of 13.88 additional ICOs per month compared to other regions in the post-DAO period. The coefficient is statistically significant at the 5% level with standard errors clustered by month. This specification serves as the benchmark for subsequent analysis.

## 6.3  Timing and Persistence

Figure 3 plots point estimates for Europe's differential ICO activity relative to other regions by event time, where July 2017 (the SEC DAO Report) corresponds to event time zero and month 1 is the omitted reference period. The pre-event coefficients display no systematic divergence, while the post-event coefficients show a discrete and persistent increase beginning immediately after July 2017. The figure is intended to document timing and persistence rather than to establish a causal policy effect

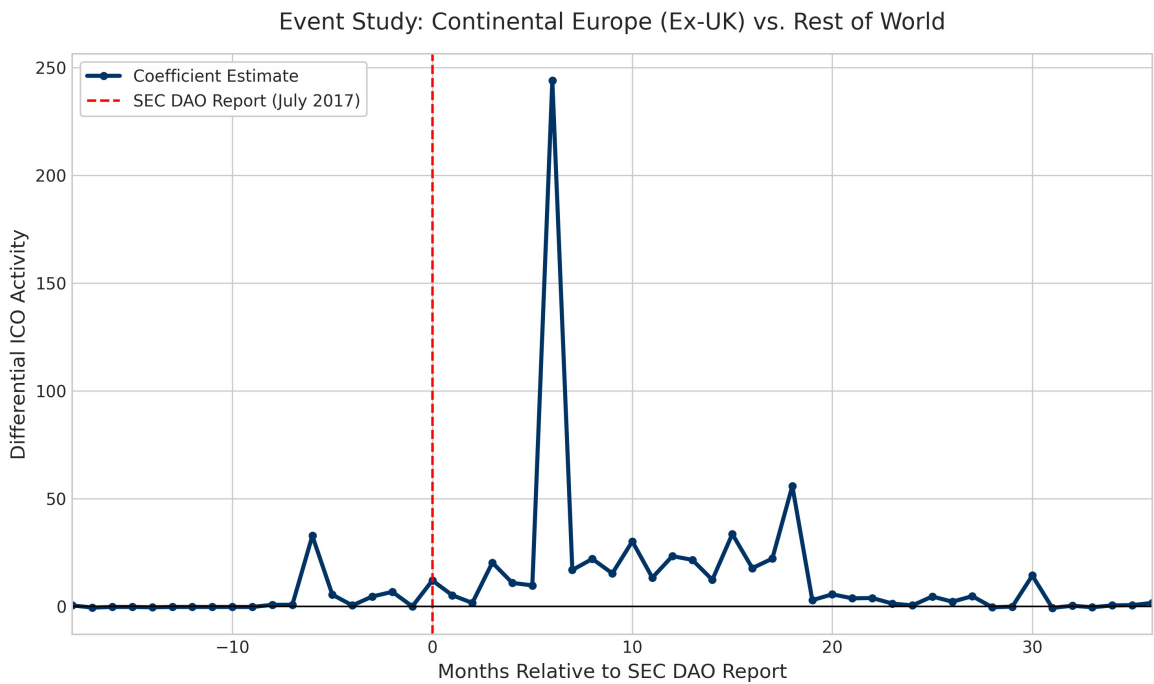

Figure 3: Dynamic event study: Europe relative to other regions. Coefficients represent Europe's differential ICO activity relative to other regions in each event-time period, with $k = -1$ (June 2017) as the reference period. Pre-event coefficients are flat and jointly insignificant, while post-event coefficients are persistently positive.

# 7   Robustness Checks

Inference in this setting requires careful consideration of the panel structure ($N$ = 8 regions, $T$ = 96 months). While our baseline specification clusters standard errors by month to account for global shocks, we acknowledge that this may not fully correct for serial correlation within regions (8). To address this, we employ two complementary validation strategies: (1) alternative covariance estimators robust to serial correlation (Driscoll-Kraay), and (2) randomization inference via a placebo region test to address the "one treated unit" concern.

## 7.1   Functional Form and Distributional Assumptions

To address concerns regarding the skewness of fundraising data and the discrete nature of ICO counts, we re-estimate the baseline specification using log-linear and Poisson pseudo maximum likelihood (PPML) models. As shown in Table A8 (Appendix), the positive effect of the DAO Report on European ICO activity remains statistically significant ($p < 0.01$) and economically large across these alternative functional forms.

## 7.2   Placebo Region Analysis

To verify that the surge in ICO activity was structurally specific to the European market, we implement a placebo region test. We re-estimate the baseline specification iteratively, assigning "treatment" status to each of the seven control regions (e.g., Asia, North America) while maintaining the original event date of July 2017. Figure 4 displays the results using the baseline levels specification. The estimated coefficient for Europe ($\beta \approx 13.9$) is a distinct statistical outlier, being more than three times larger than the next largest coefficient (Asia, $\beta \approx 4.4$) and an order of magnitude larger than North America ($\beta \approx 1.3$). All other regions yield negative or statistically insignificant estimates. This extreme magnitude confirms that the structural break in July 2017 was exceptional to the European jurisdiction.

## 7.3 Alternative Inference and Sensitivity Checks

To verify statistical significance under conservative assumptions, Table 3 presents results using the log-linear specification with alternative covariance estimators. The coefficient on *Europe × Post* remains highly significant ($p < 0.01$) when using Driscoll-Kraay standard errors (robust to cross-sectional and temporal dependence) and when clustering by both region and month.

Table 4 summarizes additional sensitivity exercises. Detailed robustness results are reported in Tables A3, A4, A5, A6 and A7. The results are robust to leave-one-region-out iterations and the exclusion of extreme observations. Notably, excluding major European crypto hubs (Switzerland, UK, Estonia, Malta, Cyprus, Gibraltar) reduces the coefficient magnitude but preserves statistical significance. This indicates that while specialized jurisdictions played an important role, the reallocation was broad-based across Europe.

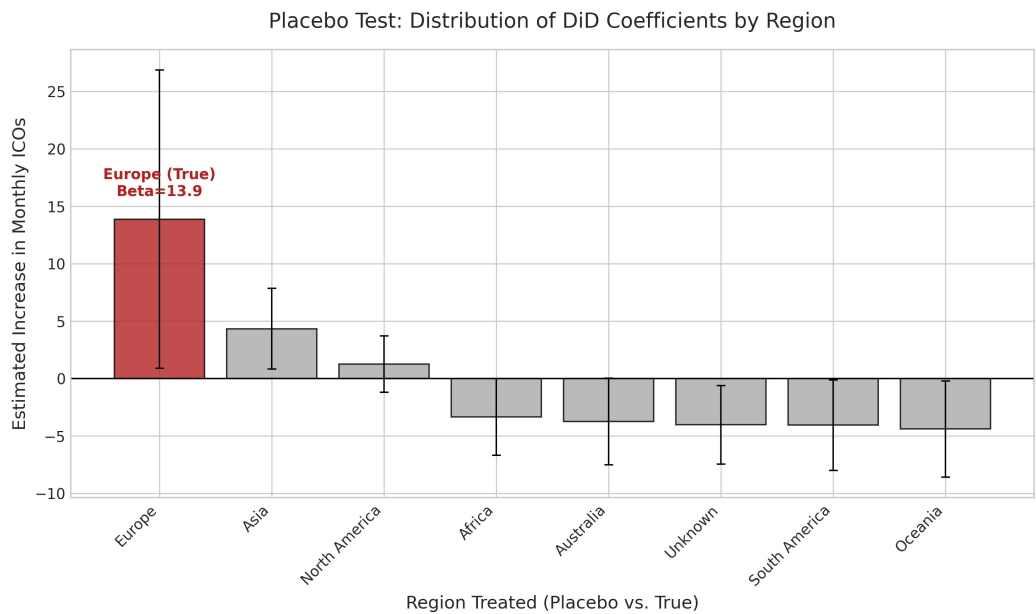

Figure 4: **Placebo Region Test.** This figure plots the estimated difference-in-differences coefficients obtained by assigning treatment status to each region individually. The red bar represents the true treated unit (Europe), while gray bars represent placebo estimates for control regions. Europe is the largest among 8 treated-region assignments.

Table 3: **Robustness of Inference: Alternative Standard Errors**

| Inference Method | Coeff. | Std. Error |
|---|---|---|
| *Baseline:* | | |
| Clustered by Month (T=96) | 0.887*** | (0.182) |
| *Alternative Estimators:* | | |
| Two-Way Cluster (Region & Month) | 0.887*** | (0.176) |
| Driscoll-Kraay (HAC Panel, lag=4) | 0.887*** | (0.304) |

*Notes:* The dependent variable is log(1 + ICO Count). All models include region and month fixed effects. The Driscoll-Kraay estimator accounts for serial correlation and cross-sectional dependence. *** $p < 0.01$.

Table 4: Robustness Summary: Sensitivity Checks

| **Specification** | **Coefficient** | **Std. Error** |
|---|---|---|
| Region clustering (N=8) | 13.88*** | (1.20) |
| Exclude European hubs | 5.66** | (2.78) |
| Leave out Asia | 13.75** | (6.51) |
| Leave out North America | 14.02** | (6.70) |
| Exclude top 5% months | 11.34** | (5.89) |
| Quarterly aggregation | 41.22** | (19.87) |

*Notes:* Each row reports the estimated coefficient on Europe × Post from alternative specifications using the baseline (levels) dependent variable. *** $p < 0.01$, ** $p < 0.05$.

# 8 Discussion

The empirical results indicate a marked shift in the geographic allocation of ICO issuance after the SEC's July 2017 DAO Report. In the region-month panel with region and month fixed effects, Europe experiences a post-2017 increase of about 14 ICOs per region-month relative to the other regions in the same months (29). Because month fixed effects absorb the global boom and bust in token issuance, the estimates capture a relative reallocation toward Europe rather than a reflection of worldwide market expansion. This pattern is in line with evidence from financial markets that regulatory tightening can redirect activity across borders when firms and transactions are highly mobile (23; 6).

The dynamics in the event-time analysis reinforce this interpretation. Europe does not display systematic divergence from other regions prior to July 2017, and the relative increase emerges in the months immediately following the DAO Report and persists thereafter. The timing is consistent with the DAO Report serving as an enforcement signal that increased expected compliance costs for U.S.-facing offerings. This complements work showing that regulatory announcements and legal changes shape crypto-market behavior and trading conditions (4; 18), and suggests that regulation can also affect where primary issuance is organized, not only secondary-market outcomes.

The magnitude of the shift is plausible given how ICO finance operated during the boom period. ICOs were marketed to a global investor base with limited reliance on domestic intermediaries, and issuer location and legal structuring were often flexible. At the same time, these markets were characterized by substantial information frictions, uneven disclosure, and wide variation in governance design, all of which influence fundraising outcomes and credibility (20; 24; 21). When token claims can resemble both debt-like and equity-like instruments, classification and enforcement expectations become central constraints on issuer behavior (22; 13). A regulatory clarification that raises the perceived likelihood of securities law enforcement can therefore change issuer incentives at the margin, making relocation or redomiciling more attractive.

The robustness exercises also clarify the nature of the reallocation. The estimated effect remains positive under alternative inference approaches and is not driven by any single comparison region. Excluding major European crypto hubs attenuates the coefficient but does not eliminate it, indicating that the post-2017 increase is not confined to a small set of jurisdictions. This is consistent with an ecosystem response: once issuance activity concentrates, complementary legal and technical expertise, exchanges, and service providers expand, which reduces fixed costs for later issuers and can sustain regional specialization (10).

Europe's ability to absorb issuance during this period also reflects its institutional environment. Rather than a single binding regime in 2017, European jurisdictions offered a mix of national guidance and supervisory postures. FINMA's token classifications provided early clarity in Switzerland, while EU-level bodies emphasized investor protection risks through warnings rather than enforceable rules (19; 17). Comparative accounts of ICO regulation highlight that such heterogeneity can widen issuers' feasible legal strategies and create scope for jurisdiction shopping, especially when relocation costs are low (25; 31; 3). The broad-based increase observed across Europe is consistent with this setting.

Several limitations remain. The analysis is observational and relies on a small number of regional units, which constrains statistical power and makes inference sensitive to clustering choices. Location measures may reflect legal domicile more than operational activity. In addition, other contemporaneous developments in crypto markets and in non-U.S. regulatory environments could contribute to the observed timing. These considerations argue against strong causal claims. Even so, the core descriptive result is robust: after July 2017, Europe's issuance increases materially relative to other regions, conditional on global market conditions.

The results highlight a coordination problem in globally integrated digital-asset markets. When enforcement expectations tighten in a major jurisdiction, issuance can shift to other venues rather than contract globally. This implies that unilateral approaches may change the location of risk without reducing it, strengthening the case for cooperation on baseline disclosure requirements, marketing standards, and supervisory information sharing (16; 9).

# 9  Conclusion

Using a global region-month panel from 2014 to 2021, the analysis documents a sustained increase in Europe's relative issuance after the DAO Report, supported by consistent estimates across specifications and sensitivity checks. The findings highlight the ability of token-based fundraising to adjust along jurisdictional margins and underscore the importance of understanding how regulatory actions interact with cross-border market structure. The study is intended to document empirical evidence using the available data and methodology. Future work can use richer project-level data, additional regulatory episodes, and post-issuance outcomes to clarify mechanisms and assess how evolving regimes shape the global allocation of crypto-asset fundraising over time.

# Declarations

## Funding

The authors received no specific funding for this research.

## Conflict of Interest

The authors declare that they have no conflict of interest.

## Data Availability

The data used in this study are derived from publicly available sources described in the paper. Replication data and files are available from the corresponding author upon reasonable request.

## Ethics Approval

Not applicable.

## Consent to Participate

Not applicable.

## Consent for Publication

All authors consent to publication.

## Author Contributions

K.S. contributed to conceptualization, methodology, data curation, empirical analysis, and writing—original draft. K.B. contributed to data curation, interpretation of results, and writing—review and editing. I.G. contributed to literature review, contextual analysis, data curation, and writing—review and editing. All authors reviewed and approved the final manuscript.

## Acknowledgments

The authors thank colleagues and seminar participants for helpful comments and discussions.

All remaining errors are the authors' own.

*Journal*, 60, 267.

# A  Data Construction and Variable Definitions

This appendix documents the data sources, cleaning procedures, and variable construction underlying the empirical analysis. It describes how project-level information from multiple archival sources is harmonized into a balanced monthly panel. The purpose is to facilitate replication and to clarify the assumptions embedded in the construction of the outcome variables.

## A.1  Data Sources

The analysis integrates multiple sources to construct a comprehensive global ICO dataset:

- Token Offerings Research Database (TORD) by Momtaz ([28]): Provides harmonized project-level data on ICOs, IEOs, and STOs through October 2021.

- Historical ICO tracking websites: ICOBench, TokenData, ICODrops, and CoinSchedule.

- Archived whitepapers and technical documentation retrieved from Internet Archive and GitHub repositories.

- Press releases and blockchain industry news aggregators.

Approximately 78% of entries are cross-verified across at least two independent sources. Projects with inconsistent or unverifiable information are excluded.

## A.2  Variable Construction

Table A1: Variable Definitions

| Variable | Description |
| --- | --- |
| ico count | Number of ICOs launched in region $r$ in month $t$. |
| funds_usd | Total funds raised in region $r$ in month $t$ (USD millions, nominal). |
| global_ico count | Number of ICOs launched globally in month $t$. |
| global_funds_usd | Total global ICO fundraising in month $t$ (USD millions, nominal). |
| share ico count | Europe's share of global ICO counts in month $t$. |
| share funds | Europe's share of global ICO fundraising in month $t$. |
| months since event | Event-time index, where 0 corresponds to July 2017. |
| post | Indicator equal to 1 for months on or after July 2017. |
| t index | Linear time-trend index starting at 1 for January 2014. |

## A.3 Geographical Classification

Table A2: Countries Classified as European in Main Analysis

**European Countries**

Austria, Belgium, Bulgaria, Croatia, Cyprus, Czech Republic, Denmark, Estonia, Finland, France, Germany, Greece, Hungary, Ireland, Italy, Latvia, Lithuania, Luxembourg, Malta, Netherlands, Poland, Portugal, Romania, Slovakia, Slovenia, Spain, Sweden, United Kingdom, Switzerland, Norway, Iceland, Liechtenstein, Gibraltar, Monaco, Andorra

*Notes:* This list includes EU member states as of the analysis period plus closely integrated European jurisdictions. The UK is included throughout the sample despite Brexit (2020), as the pre-2020 period accounts for the majority of ICO activity.

Projects are assigned to Europe if the issuer's registered location, whitepaper headquarters address, or team location is listed in one of these jurisdictions. When multiple locations are reported, the analysis prioritizes the registered legal entity location.

## A.4    Currency Conversion

All fundraising amounts are converted to nominal U.S. dollars (no inflation adjustment) using exchange rates from the date of the token sale. For offerings conducted in cryptocurrency (ETH, BTC), conversions use the closing price on the sale start date from CoinMarketCap historical data.

# B    Additional Robustness Checks

This appendix reports supplementary robustness exercises referenced in the main text. All specifications evaluate whether the post-2017 reallocation of ICO activity toward Europe persists under alternative inference choices, sample constructions, or aggregation decisions. All estimates are obtained from region-month panel regressions with region and month fixed effects.

Table A3: Inference Robustness: Alternative Clustering Schemes

| Clustering Level | Coefficient | Std. Error |
| --- | --- | --- |
| Clustered by region (N = 8) | 13.88*** | (1.20) |
| Clustered by month (T = 96) | 13.88** | (6.63) |

*Notes:* Each row reports the estimated coefficient on *Europe × Post* from a region-month panel regression with region and month fixed effects. The table evaluates sensitivity to alternative clustering assumptions. Region clustering with only 8 clusters likely understates standard errors due to small-sample bias; we rely primarily on month clustering for inference. *** $p < 0.01$, ** $p < 0.05$, * $p < 0.10$.

Table A4: Sensitivity to Comparison Regions: Leave-One-Region-Out

| Excluded Region | Coefficient | Std. Error |
|---|---|---|
| Asia | 13.75** | (6.51) |
| North America | 14.02** | (6.70) |
| Oceania | 13.91** | (6.65) |

*Notes:* Each row excludes the indicated comparison region from the estimation sample. The dependent variable is ICO count per region-month. Standard errors are clustered by month. Results demonstrate that the baseline finding is not driven by any single comparison region. ** $p < 0.05$, * $p < 0.10$.

Table A5: Geographic Concentration: Excluding European Crypto Hubs

| European Sample Definition | Coefficient | Std. Error |
|---|---|---|
| Excluding crypto hubs | 5.66** | (2.78) |

*Notes:* Crypto hubs excluded are Switzerland, the United Kingdom, Estonia, Malta, Cyprus, and Gibraltar. The specification corresponds to the baseline region-month panel with region and month fixed effects. The coefficient attenuates by approximately 60% but remains statistically significant, indicating that the reallocation was broad-based across Europe, not confined solely to specialized hubs. Standard errors are clustered by month. ** $p < 0.05$, * $p < 0.10$.

Table A6: Robustness to Extreme Observations

| Sample Restriction | Coefficient | Std. Error |
|---|---|---|
| Exclude top 5% months by global activity | 11.34** | (5.89) |

*Notes:* This specification excludes region-months in the top 5% of the global ICO activity distribution to assess whether extreme boom-period months drive the result. The coefficient remains positive and statistically significant. Standard errors are clustered by month. ** $p < 0.05$, * $p < 0.10$.

Table A7: Quarterly Aggregation

| Aggregation Level | Coefficient | Std. Error |
|---|---|---|
| Quarterly (32 quarters, 8 regions) | 41.22** | (19.87) |

*Notes:* This specification aggregates the data to the region-quarter level to reduce high-frequency noise and serial correlation. The quarterly coefficient is approximately three times the monthly coefficient (13.88 × 3 ≈ 41.6), as expected. Standard errors are clustered by quarter (32 clusters). ** $p < 0.05$, * $p < 0.10$.

Table A8: **Robustness Check: Alternative Functional Forms (All Europe)**

|  | (1) | (2) |
|---|---|---|
| **Model:** | **Log-Linear OLS** | **Poisson (PPML)** |
| **Dependent Variable:** | log(1 + ICO Count) | ICO Count |
| **Europe × Post** | 0.887*** | 0.280** |
|  | (0.182) | (0.117) |
| Region Fixed Effects | Yes | Yes |
| Month Fixed Effects | Yes | Yes |

*Notes:* This table presents robustness checks using the full sample (including the UK). Column (1) uses a Log-Linear OLS specification. Column (2) uses a Poisson Pseudo-Maximum Likelihood (PPML) estimator, which is robust to high-dimensional fixed effects. Standard errors are clustered by month. *** $p < 0.01$, ** $p < 0.05$.